\documentclass[apjl]{emulateapj}
\usepackage[colorlinks=True,linkcolor=blue,citecolor=blue,urlcolor=blue]{hyperref}    
\usepackage{graphicx}
\usepackage{amsmath}
\usepackage{natbib}
\usepackage[table]{xcolor}
\usepackage{multirow}
\usepackage{rotating}
\usepackage{tabu}

\usepackage[utf8x]{inputenc}
\usepackage[table]{xcolor}
\usepackage{tabu}
\usepackage{ulem}

\newcommand\degree{$^{\circ}$}

\newcommand\simlt{\lower.5ex\hbox{$\; \buildrel < \over \sim \;$}}
\newcommand\simgt{\lower.5ex\hbox{$\; \buildrel > \over \sim \;$}}

\begin{document}
\title{The extragalactic ultra-high energy cosmic-ray dipole}

\author{Noemie Globus\altaffilmark{1} and Tsvi Piran\altaffilmark{1}}
\altaffiltext{1}{Racah Institute of Physics, The Hebrew University, 91904 Jerusalem, Israel}

\begin{abstract}
We explore the possibility that the recently detected dipole anisotropy in the arrival directions of~$>8$~EeV ultra-high energy cosmic-rays (UHECRs) arises due to the large-scale structure (LSS).  We assume that the cosmic ray sources follow the matter distribution and calculate the flux-weighted UHECRs' RMS  dipole amplitude taking into account the diffusive transport in the intergalactic magnetic field (IGMF).  We find that the flux-weighted RMS dipole amplitude is $\sim8$\% before entering the Galaxy. The amplitude in the [4-8] EeV is only slightly lower $\sim 5$\%. The required IGMF is of the order of {5-30 nG}, and the UHECR sources must be relatively nearby, within $\sim$300 Mpc. { The absence of statistically significant signal in the lower energy bin can be explained if the same nuclei specie dominates the composition in both energy bins and diffusion in the Galactic magnetic field (GMF) reduces the dipole of these lower rigidity particles.  Photodisintegration of higher energy UHECRs could also reduce somewhat the lower energy dipole. }
\end{abstract}
\keywords{cosmic rays}

\section{Introduction}
The origin and nature of ultra-high energy cosmic-rays (UHECRs) are still open questions.
The  Pierre Auger Observatory  has recently reported   a significant ($>5\sigma$) anisotropy in the arrival directions of UHECRs \citep{Taborda17}.  
It is a large scale hemispherical asymmetry in the cosmic-rays arrival directions,  
well-represented by a dipole with an amplitude  $A =(6.5^{+1.3}_{-0.9})$\% pointing in the direction  
 ($l$, $b$) = (233\degree, -13\degree) $\pm10$\degree$\,$  in Galactic coordinates.
The dipole anisotropy is detected above 8 EeV.
{ In the energy bin [4-8] EeV, the amplitude of the dipole is $A =(2.5^{+1.0}_{-0.7})$\%. However this amplitude has low statistical significance, and  the data in this energy bin are compatible with isotropy   \citep{Taborda17}.} The median energies for these two bins are 5.0 EeV and 11.5 EeV, respectively.
 
We explore, here,  the possibility that the observed dipole arises due to a large-scale structure (LSS) anisotropy.  
If the UHECR sources follow the LSS, different UHECR energies would probe different distances, 
as a result of the energy dependent GZK horizon. With enough UHECR events, we would obtain a tomographic view of the LSS \citep{Waxman+97},  in a similar manner to 
 intensity mapping  \citep{Madau97}. Following \citet{Lahav+97}, we calculate the expected RMS dipole  in the observed  cosmic-ray intensity that will arise due to fluctuations in the source distribution, assuming that those follow the LSS.  This is a measure of the expected dipole and also of the fluctuations of this value. Since  UHECR diffuse in the Inter Galactic Magnetic Field (IGMF) we add magnetic diffusion into the model.  We expand the surface brightness of the cosmic-ray sky  in spherical harmonics of order $l$, where $l=1$ corresponds to the dipole. We express these harmonics in terms of the power spectrum $P(k)$ of the matter distribution  and calculate the  dipole amplitude for different values of the IGMF. 

We make two  approximations. First, we neglect fluctuations due to individual sources.  This is justified given the large number of observed UHECRs in these energy ranges and by the absence of strong small scale anisotropies. We are interested in the dipole, that corresponds to large scales - that are dominated by the distribution as a whole.  Second, we neglect nuclei photodisintegration. A quantitive analysis of this effect  requires further assumptions on the specific composition, which is unknown. Instead we  present a general model and show in   \S \ref{sec:discussion} that this assumption doesn't affect significantly the  the anisotropy at the high energy band and  it decreases somewhat the anisotropy at  the lower energy range.

The structure of the paper is as follows. 
We begin in \S~\ref{sec:qualitative} with a qualitative description of the model. 
Our analysis depends on the composition of the UHECRs and hence we briefly  summarize in \S~\ref{sec:Composition} the interpreted Auger composition.   
We discuss in \S~\ref{sec:horizons} the notion of the cosmic-rays horizon, {\it i.e.} the fraction of the Universe that would contribute to the observed anisotropy. 
We describe the methods in \S~\ref{sec:methods}. We present the {dipole that arises from the LSS in \S~\ref{sec:results} 
and discuss the  interpretation of the results in \S~\ref{sec:discussion}.

\section{Qualitative discussion} 
\label{sec:qualitative}
The idea to use UHECRs to probe the LSS was proposed by \citet{Waxman+97}.  If the  IGMF is negligible, the UHECR horizon is the GZK distance. At energies at which the dipole is observed, this distance is almost the Hubble distance, and therefore, the UHECR dipole axis should be aligned with the CMB dipole (that follows the LSS dipole matter distribution).  However, both the amplitude and the direction of the UHECR dipole are not compatible with  the CMB dipole.

With a significant IGMF, the UHECR horizon is the magnetic horizon defined in \S \ref{sec:horizons}. This horizon  is smaller than the GZK distance. It decreases  
 with increasing IGMF and with decreasing UHECR rigidity. 
A smaller horizon leads to  larger LSS fluctuations \citep[e.g.][]{Lahav+97} and thus it enhances the anisotropy.
At the same time, the diffusion in the stronger IGMF weakens the anisotropy. 
The first effect dominates in the quasi-rectilinear regime while the second dominates in the diffusive regime.

In addition to diffusion in the IGMF the UHECRs, at these energies,  are both deflected and scattered in the Galactic magnetic field (GMF). This  may change both the magnitude and the  direction of the dipole.

The dipole anisotropy has been observed at  $>$8~EeV, but  not  in the [4-8] EeV energy bin.
If the same species dominates  both energy bins, then a stronger diffusion in the magnetic fields would lead to a lower dipole amplitude in the lower energy bin. If different species dominate and  we have a similar rigidity in both energy bins and hence, unless the effective magnetic horizons are very different, we expect comparable dipole amplitudes.

 \section{Composition}
\label{sec:Composition}

At $>8$ EeV, where the dipole is reported\footnote{Note that most of the UHECRs in this energy range have $E<30$ EeV.},   the Auger composition data \citep{Bellido17} suggests that there are  less than 20\% protons.
While this fraction is very similar for the three hadronic models used to reconstruct the composition,  the exact nature of the dominant nuclei specie at {these energies}  is still debated. All models indicate that the composition becomes heavier with increasing energy.
When reconstructed using the EPOS-LHC or Sybill 2.3 models, the Auger data  indicates a mixture of He and CNO elements at [8-30] EeV.  In these models, CNO dominate at 30 EeV and  the transition seems to arise at $\sim 15$ EeV, but this has large uncertainty.
When reconstructed using the QGSJETII 04 model, the composition seems to be dominated by He up to $\sim30$ EeV. In the energy bin [4-8] EeV, where the Auger skymap is compatible with isotropy, the composition is lighter. It is dominated by protons (at least $\sim$60\% in the case of QJSJETII 04) or He (when EPOS-LHC or Sybill 2.3 are used).

\section{cosmic-ray horizons}
\label{sec:horizons}
Above a certain threshold energy, the interactions between cosmic rays and the extragalactic photon backgrounds limits the distances that an UHECR can travel, known as the GZK distance. At around 10 EeV, the energy at which  
the cosmic-ray dipole anisotropy is observed,  the GZK distance, $d_{\rm GZK}$, is  $\sim 1000$ Mpc for He and $\sim 2000$ Mpc for protons and CNO. At 4 EeV, the  GZK distance is $\sim 2000 $ Mpc for protons and the Hubble distance for He and CNO (see Fig.~\ref{dGZK}).
\begin{figure}[!h]
       \centering
	\includegraphics[scale=0.45]{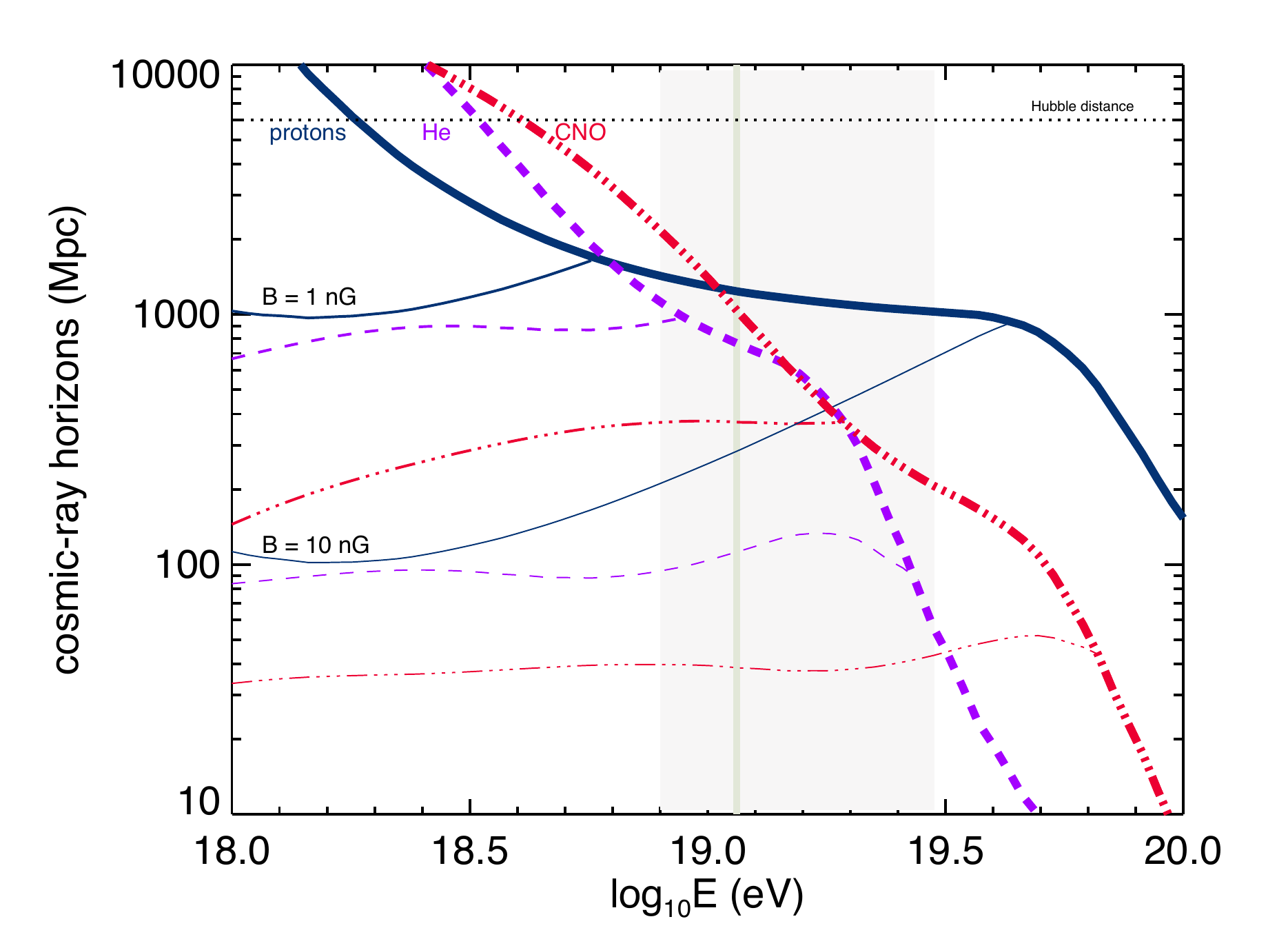}			
		\caption{Thick lines: The GZK distance $d_{\rm GZK}$ for protons (solid line), He (violet dashed line) and CNO (red dotted dashed line). Thin lines: the diffusion distance for protons, He and CNO for  1 and 10 nG IGMF and a field coherence length of 0.2 Mpc. The horizon is $H=\min(\sqrt{6D\tau_{\rm GZK}},d_{\rm GZK})$. The shaded area indicates the energy range at which the dipole anisotropy has been observed. The median value of the energy bin (11.5 EeV) is marked by a solid line.}
		\label{dGZK}
\end{figure}

While the lifetime of an UHECR is limited to $d_{\rm GZK}/c$, diffusion in the IGMF 
limits its magnetic horizon to much less than the GZK distance. 
For a diffusive propagation, the horizon scale is  the diffusion distance $\sim~\sqrt{6D \min(\tau_{\rm GZK}, t_{\rm age})}$,  where $\tau_{\rm GZK}$ is the GZK time, $t_{\rm age}$  is the age of the source 
{(the Hubble time if the source is always active)}.  The diffusion coefficient, $D$, depends on the rigidity of the particles and on the 
{strength} 
 and coherence length  of the magnetic field. For a Kolmogorov turbulence, 
  $D$ is well approximated by a fitting function taking into account both  the resonant and non-resonant diffusion regimes \citep{GAP08},
\begin{equation}
D\approx0.03\left(\frac{\lambda_{\rm Mpc}^2 E_{\rm EeV}}{ZB_{\rm nG}}\right)^{\frac{1}{3}}+0.5\left(\frac{E_{\rm EeV}}{ZB_{\rm nG}\lambda_{\rm Mpc}^{0.5}}\right)^{2} {\rm Mpc^2 Myr^{-1}}
\label{dcoef}
\end{equation}
where $Z$ is the charge of the cosmic-ray,   $E_{\rm EeV}$ is its energy measured in EeV, 
$B_{\rm nG}$ the IGMF strength in nG and $\lambda_{\rm Mpc}$ its coherence length in Mpc.   
The diffusive approximation typically holds for ${6D<d_{\rm GZK}c}$, {\it i.e.} when the diffusion distance $\sqrt{6D\tau_{\rm GZK}}$ is smaller than the GZK distance.
When it takes more than the age of the Universe to enter the diffusion regime, then the cosmic-ray propagation is (quasi-) rectilinear.
The size of the region that  contributes to the observed flux and hence to the anisotropy is thus set by the cosmic-ray horizon  \citep{2004NuPhS.136..169P}
\begin{equation}
H=\min(\sqrt{6D\tau_{\rm GZK}},d_{\rm GZK}).
\label{horizon}
\end{equation}
 Fig.~\ref{dGZK} depicts $d_{\rm GZK}$ and  $H$ for two different values of the IGMF  $B_{\rm nG}=1$ and $B_{\rm nG}=10$. At the energies at which the dipole anisotropy has been observed ($8< E_{\rm EeV}\lesssim30$, see the shaded area on Fig. \ref{dGZK}) the diffusion approximation holds for  $B_{\rm nG}\gtrsim10$.

\section{Spherical harmonic expansion of background sources}
\label{sec:methods}

 \begin{figure}[!t]
   \centering
	\includegraphics[scale=0.25]{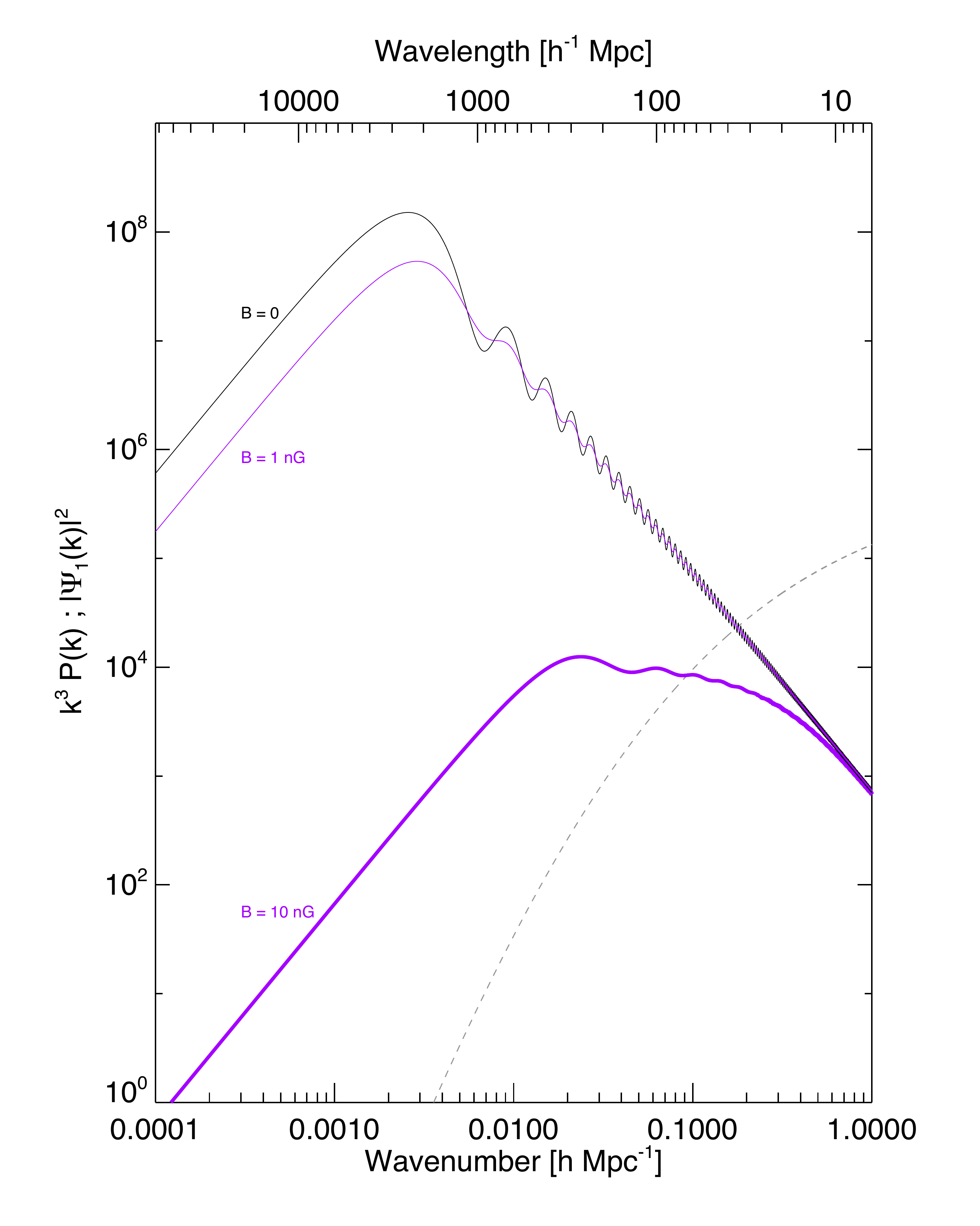}				
		\caption{The dipole window functions $|\Psi_{l=1}(k)|^2$, where ${<|a_{lm}|^2>}\varpropto\int dk k^2 P(k) |\Psi_l(k)|^2$, as given by Eq. \ref{window} in the text, here calculated for He at 11.5 EeV for  $B_{\rm nG}=0$, 1 and 10. 
		The functions weighs the contribution from each plane wave $k$ to the dipole $l=1$. The dashed line represents $k^3P(k)$ for a standard $\Lambda$CDM cosmological model. 
		The vertical scaling is arbitrary.}
		\label{window_function}
\end{figure}

To calculate the UHECR anisotropy due to LSS, we modify the formalism introduced by \citet{Lahav+97} to include diffusion in the IGMF. 
We expand the surface brightness of the cosmic-ray sky in spherical harmonics,
\begin{equation}
I(\hat{e})=\sum_{lm}a_{lm}Y_{lm}(\hat{e}).
\end{equation}
Due of the diffusion in the IGMF, the cosmic-ray anisotropy of a shell at a radius $r$ is reduced by a factor $\alpha_{lm}(r)$ relative to the sources' distribution.    We write 
the harmonic coefficients, $a_{lm}$, as:
\begin{equation}
a_{lm}=\frac{1}{4\pi}\rho_0\int \int d\Omega dr \delta(r,\hat{e})\alpha_{lm}(r)Y_{lm}^*(\hat{e})\,,
\label{a_lm}
\end{equation} 
 where $\delta(r,\hat{e})$ is the 3 dimensional  density contrast, expanded  using Rayleigh expansion into plane waves \citep[cf.][for details]{Lahav+97},
 \begin{equation}
\delta(r,\hat{e})=\frac{1}{(2\pi)^3}\int d^3k \delta_k\,{\rm e}^{-i\hat{k}\cdot\hat{r}} .
 \label{eq:deltar}
 \end{equation}
 
 
Plugging Eq. \ref{eq:deltar} into Eq. \ref{a_lm} we obtain:
\begin{equation}
a_{lm}=(i^l)^*\frac{1}{2\pi^2}\frac{1}{4\pi}\rho_0\int d^3k \delta_k Y_{lm}^*(\hat{k})\int_{0}^{H}dr j_l(k r) \alpha_{lm}(r)\,,
\end{equation} 
where $j_l$ is the Bessel function of order $l$, $\rho_0$ the cosmic-ray emissivity at redshift $z=0$ (in erg Mpc$^{-3}$ yr$^{-1}$) and $H$ is the cosmic-ray horizon defined by Eq. \ref{horizon}. 

The Bessel function, $j_l$, projects a wave (with a certain direction $\hat{e}$, at a certain distance $r$) onto the celestial sphere,  {\it i.e.} it relates the contribution of  each wavenumber $k$ along the line of sight  to the pattern of anisotropies at angular scale $\theta\simeq\pi/l$.

As the nature of UHECR sources is unknown, it is not clear how do they trace the matter distribution.  {Our basic assumption is that  the UHECR sources trace the matter density. However the fluctuations might be biased and thus the UHECR }
density fluctuations are  proportional to the mass density fluctuations, $\delta(r,\hat{e})$,
with a proportionally (bias) factor $b$
  ($b >1$ implies that the UHECR sources are more clustered than the mass distribution). 
The power spectrum $P(k)$ is given in terms of the current mass density fluctuations $\delta_k$ as
$<\delta_{\hat{k}} \delta^{*}_{\hat{k}'}>=(2\pi)^3 P(k)\delta^{(3)}(\hat{k}-\hat{k}')$,
where $\delta^{(3)}$ is the 3 dimensional  delta function. 
Hence, assuming constant bias, 
the  fluctuations of different multipoles are given by
\begin{equation}
<|a_{lm}|^2>=\frac{1}{(2\pi)^3}\rho_0^2b^2\int dk k^2 P(k) |\Psi_{lm}(k)|^2
\end{equation} 
where $\Psi_{lm}(k)$ is the window function, 
\begin{equation}
\Psi_{lm}(k) \equiv \int_{0}^{H}dr j_l(k r) \alpha_{lm}(r)\,.
\label{window}
\end{equation}

 As we assume that the IGMF is purely turbulent and spatially homogeneous on large scales, $\alpha_{lm}=\alpha_{lm}(r)$, {\it i.e.} is a function of $r$ only. 
 For an isotropic diffusion, that we consider here,  $\alpha_{lm}$ does not depend on $m$.  
Moreover, we are interested only in the dipole anisotropy, so we will need only $\alpha_1$.

We turn now to estimate the diffusion factor $\alpha_1$ and the amplitude of the dipole anisotropy $\Delta$.  
The cosmic-ray surface brightness 
in any direction of the sky $\hat{e}$, is  given by $I(\hat{e})=\bar{I}(1+\Delta\hat{e}\cdot\hat{\mu})$, where $\hat{\mu}$ is the unit vector pointing in the direction of the dipole and $\bar{I}$ the average intensity. 

Consider, first, a single source {from which the cosmic-rays propagate with a diffusion coefficient, $D$}. The flux  at a given distance $r$ from the source is given by $F=-D{{\partial}n}/{{\partial}r}$ (Fick's law).
The observed surface brightness {at a time $t$} as a function of angle $\theta$ to the source is $I(\theta,r,t)=(1+{\rm d}\cos\theta)\bar{I}$
where $\bar{I}={n(r,t,E)c}/{(4\pi)}$. 
The flux $F$ is obtained by integrating $I(\theta,r,t)$ over all angles \citep{Ginzburg64},
\begin{equation}
F=2\pi\int_0^\pi \bar{I}(1+{\rm d}\cos\theta) \cos\theta \sin\theta d\theta=\frac{4\pi}{3}{\rm d} \bar{I}\equiv-D\frac{{\partial}n}{{\partial}r}\,.
\label{flux}
\end{equation}
The amplitude of the dipole anisotropy due to a single source is  
\begin{equation}
{\rm d}=\frac{3D}{nc}\left|\frac{{\partial}n}{{\partial}r}\right|\equiv3\alpha_1\,,
\end{equation}
and the surface brightness, $I(\theta,r,t)=(1+3\alpha_1 \cos\theta)\bar{I}$.

We can determine $\alpha_1$ from Eq. \ref{flux}. The density $n$ of cosmic-rays propagating from an {\it instantaneous} source occurring at $-t$ in the past, and located at a distance $r$, is $n(r,t,E)=n_0(4\pi Dt)^{-3/2}\exp[-r^2/(4Dt)]$; the dipole anisotropy is  
$\alpha_1=r/(2ct)$.
For a {\it  continuous} source active over a time $t_{\rm min}$, $n(r,E)= n_0(4\pi rD)^{-1}{\rm erfc}(r/\sqrt{4D t_{\rm min}})$ where $t_{\rm min}=\min(\tau_{\rm GZK}, t_{\rm age})$. In that case  the dipole anisotropy is 
\begin{equation}
\alpha_1=\frac{D}{rc}+\frac{\sqrt{D}\exp\left(-r^2/4Dt_{\rm min}\right)}{c\sqrt{\pi t_{\rm min}}\,{\rm erfc}\left(\frac{r}{\sqrt{4 D t_{\rm min}}}\right)}\,.
\label{alphadif}
\end{equation}

In the following we consider  sources that are active over the GZK time, $t_{\rm min}\equiv\tau_{\rm GZK}$.
However, the diffusion approximation is valid only when $D<rc$ and it would be inaccurate to use Eq. \ref{alphadif} for $B_{\rm nG}\lesssim10$ at the energies of interest, as already mentioned in Sect.~\ref{sec:horizons}. 
To take into account the quasi-linear propagation, we use the result of \citet{Harari14}. They performed numerical simulations of the propagation of cosmic-rays in purely turbulent magnetic fields and derived
a fitting function $\alpha_1$ that takes into account the transition from the diffusive to the quasi-rectilinear regime, 
\begin{equation}
\alpha_1\approx\frac{D}{rc}\left[1-\exp\left(-\frac{rc}{D}-\frac{7}{18}\frac{(rc)^2}{D^2}\right)\right]\,.     
\label{alphagood}
\end{equation}

  \begin{figure}[t]
   \centering
	\includegraphics[scale=0.6]{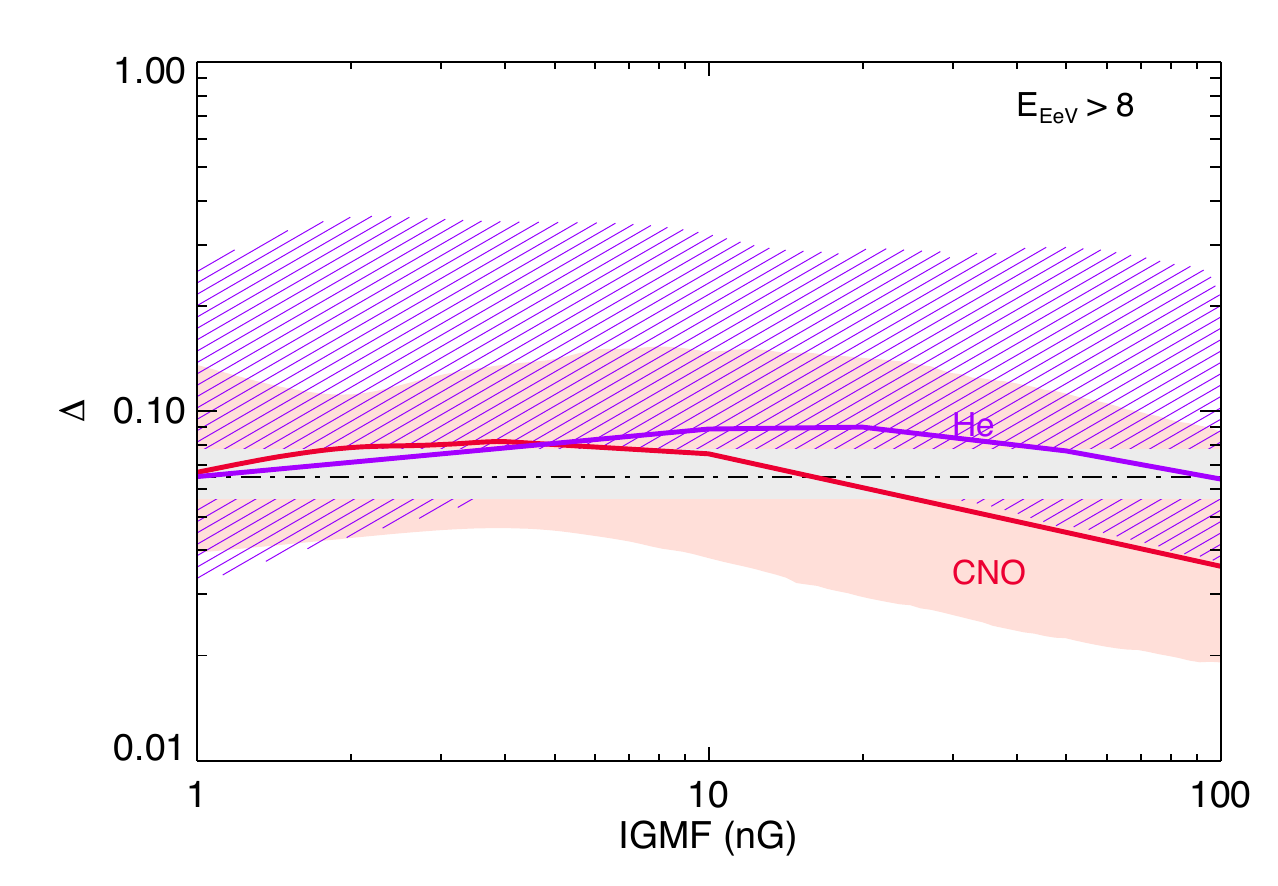} \includegraphics[scale=0.6]{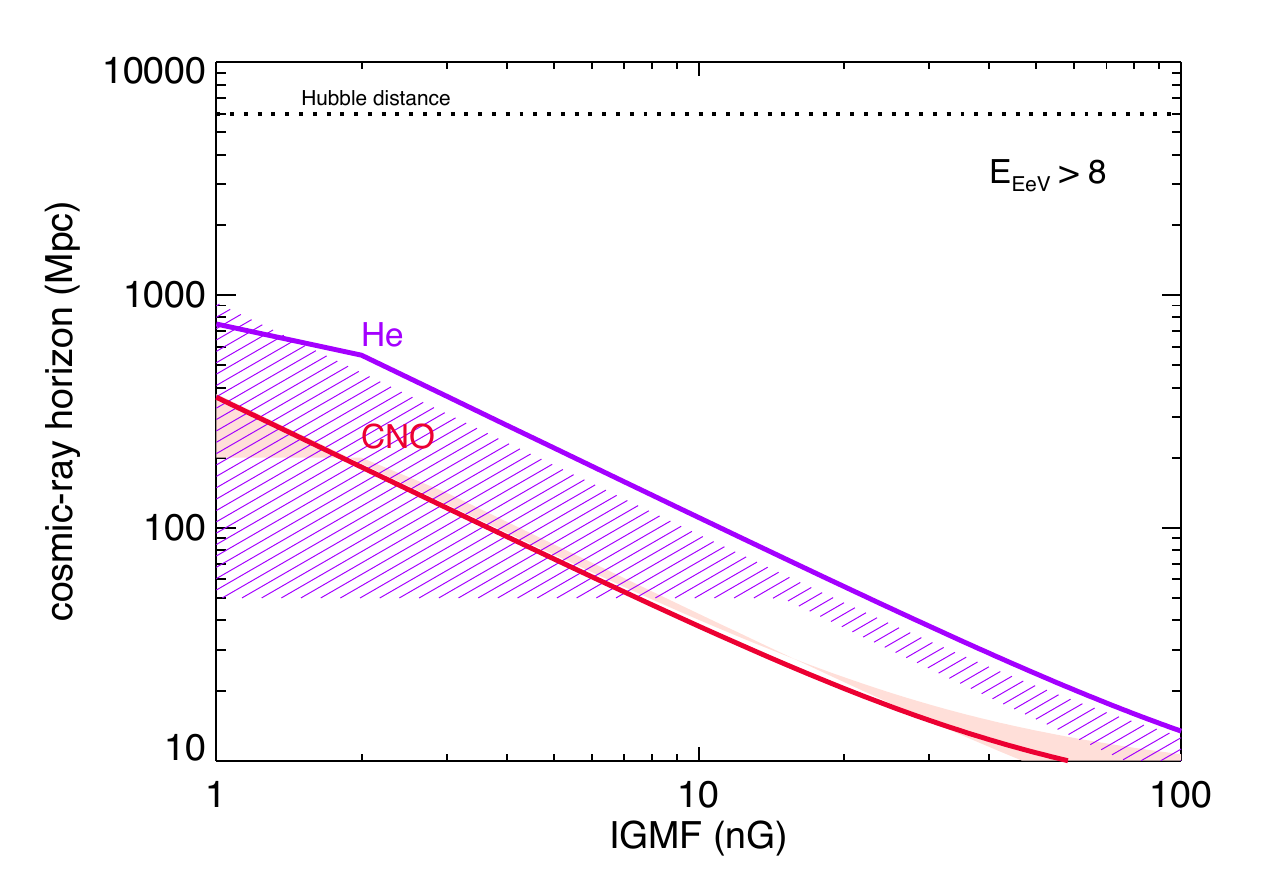}
		\caption{Upper panel: The range of amplitudes of the  dipole $\Delta$ (for $b=1$), as defined in Eq.~\ref{dipole}, 
		for different  energies in the range  [8-30] EeV for He (violet) and CNO (red). The lower limit of the shaded area corresponds to 8 EeV, the upper limit to 30 EeV. { The solid line indicate the flux-weighted  dipole.} 
                  The black dashed-dotted line indicates the observed amplitude $6.5$\% and the shaded grey area its uncertainty. Lower panel: The magnetic horizon, $H$, as given by Eq. \ref{horizon}. 
		In the case of He, the cosmic-ray horizon at 30 EeV (lower limit of the shaded area) is significantly reduced by the GZK effect.  }
		\label{amplitude_dipole8-15}
\end{figure}

\begin{figure}[t]
   \centering
	\includegraphics[scale=0.6]{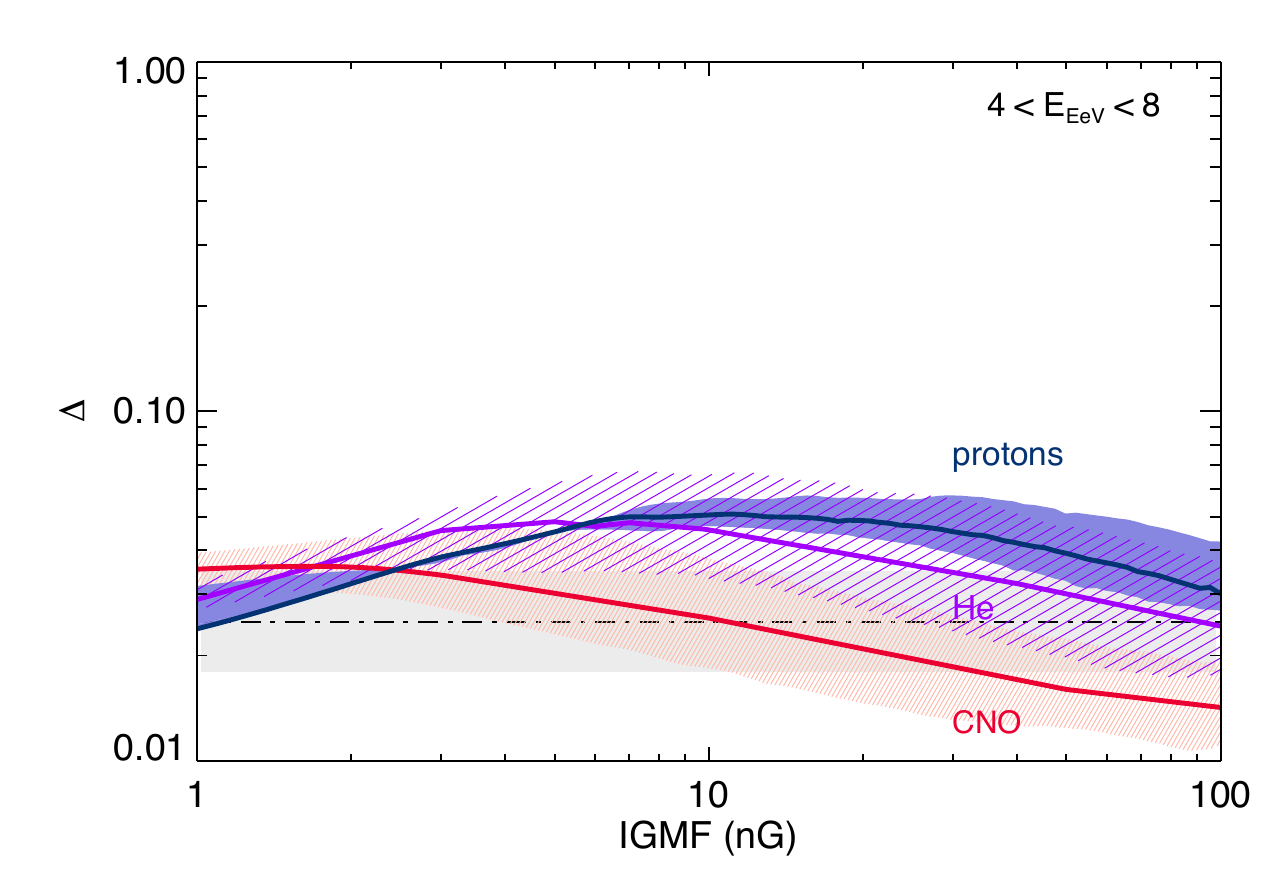}	\includegraphics[scale=0.6]{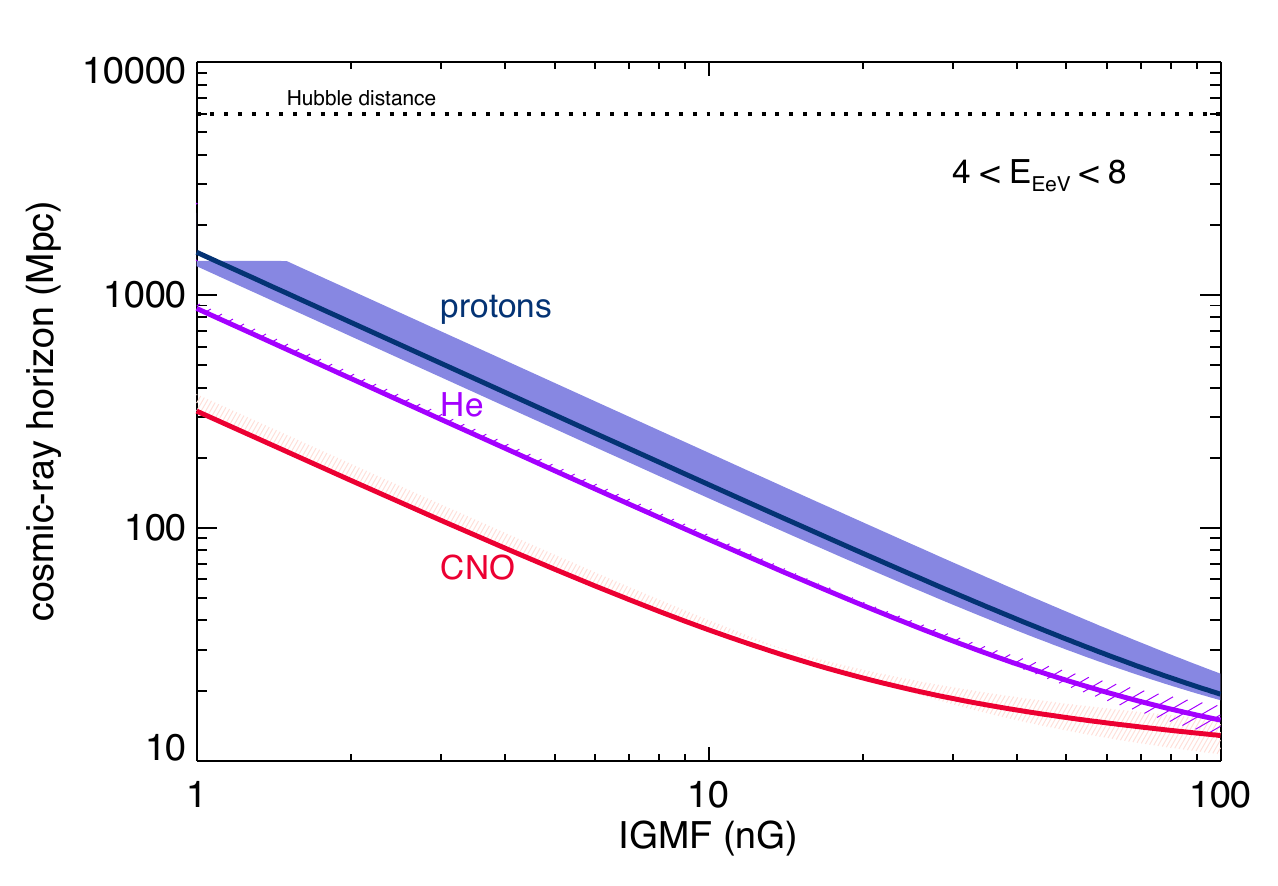}
		\caption{Upper panel: The range of amplitudes of the  dipole $\Delta$ (for $b=1$), as defined in Eq.~\ref{dipole}, for different  energies in the range  [4-8] EeV for protons (blue), He (violet), and CNO (red). 
		The lower limit of the shaded area corresponds to 4 EeV, the upper limit to 8 EeV. { The solid line indicate the flux-weighted  dipole.} 
		{ The black dashed-dotted line indicates the observed amplitude $2.5$\% (although this energy bin is also compatible with isotropy) and the shaded grey area its uncertainty.}
		Lower panel: The magnetic horizon, $H$, as given by Eq. \ref{horizon}. }
		\label{amplitude_dipole4-8}
\end{figure}

Turning now to multiple sources, the dipole anisotropy is  $\hat{\Delta}=\sum  {\rm d}_i\hat{e}_i$ where ${\rm d}_i=3\alpha_{1,i}$. The sum of the contribution of  numerous  discrete sources can be approximated by integration over a  continuous source density field.  Applying the above formalism, 
the average UHECR surface brightness is $\bar{I}=a_{00}/\sqrt{4\pi}={\rho_0}H /(4\pi)$,  and the  dipolar fluctuation in the UHECR surface brightness is  
\begin{eqnarray}
\sqrt{<\delta I^2>}&&=\sqrt{\frac{3}{4\pi}\left(\sum_{m=-1}^{1}<|a_{1m}|^2>\right)}
\nonumber\\&&
\equiv\frac{3}{\sqrt{4\pi}}\sqrt{<|a_{1}|^2>} \ , 
\end{eqnarray}
{where $a_1=a_{1m}$ and we have used the fact that the $a_{lm}$ coefficients are independent of $m$.}
Overall, the total  
dipole $\Delta$ is:
\begin{equation}
\Delta=\frac{3<|a_{1}|^2>^{1/2}}{a_{00}}\,.
\label{dipole}
\end{equation}

To visualize the scales probed by the UHECR background, we show in Fig.~\ref{window_function} the dipole  window functions $|\Psi_{l=1}(k)|^2$ for different values of the IGMF.

  \begin{figure}[t]
   \centering
	\includegraphics[scale=0.6]{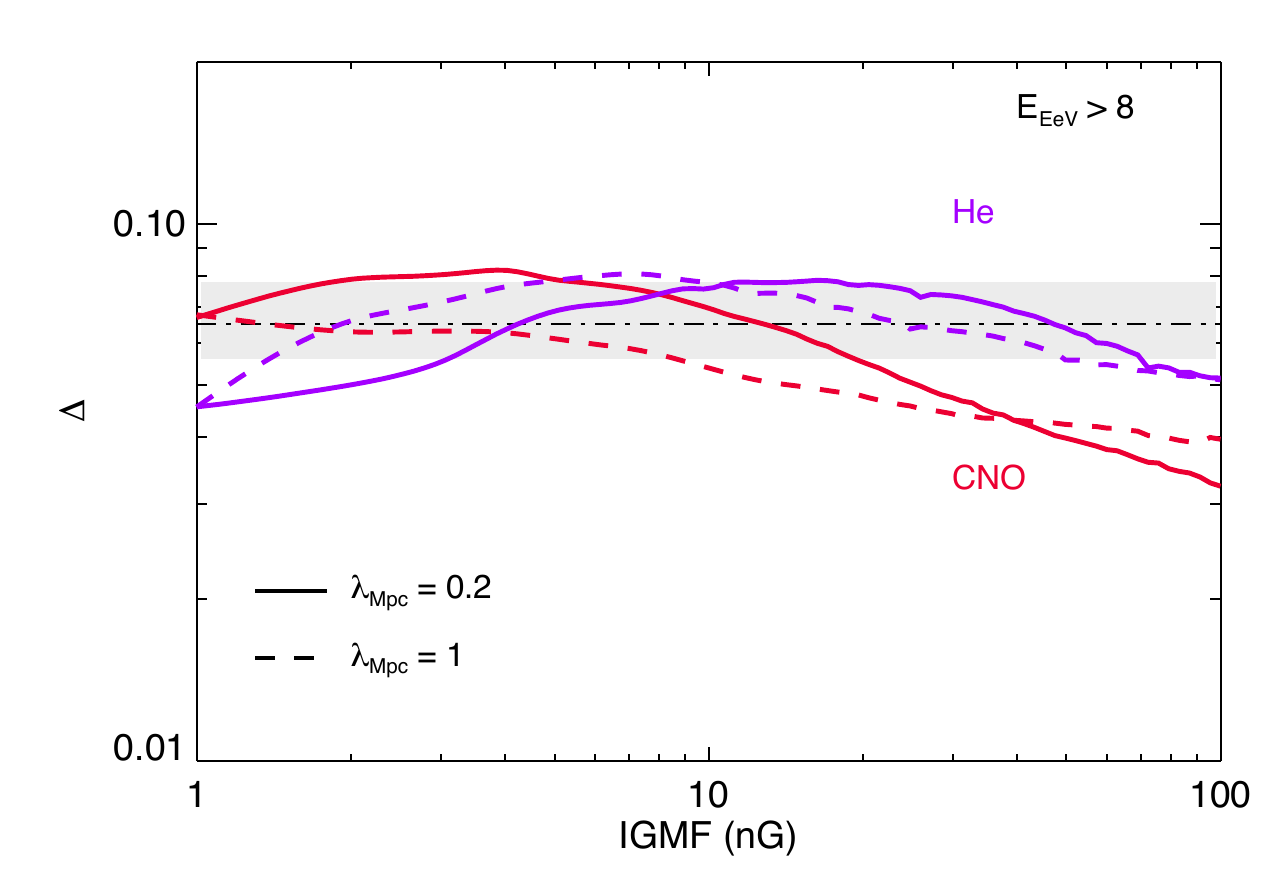} \includegraphics[scale=0.62]{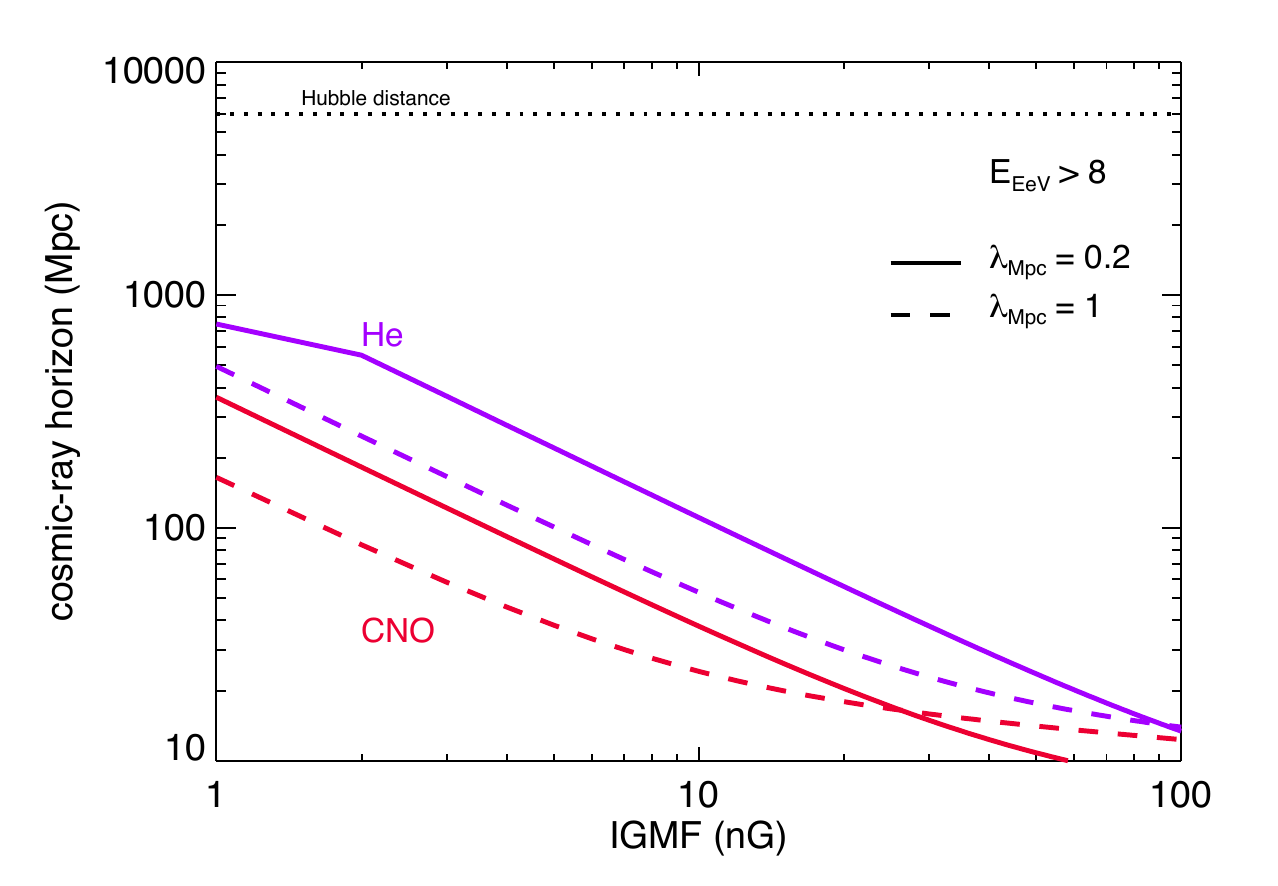}		
\caption{ The effect of varying $\lambda_{\rm Mpc}$. Upper panel: The  dipole amplitude $\Delta$ (for $b=1$), { calculated for the median value of the energy bin, $\Delta (E_{\rm EeV}=11.5)$}, 
for $\lambda_{\rm Mpc}=0.2$ (solid lines) and $\lambda_{\rm Mpc}=1$ (dashed lines).   The black dashed-dotted line indicates the observed dipole amplitude $6.5$\% and the shaded grey area its uncertainty. Lower panel: The magnetic horizon, $H$, as given by Eq. \ref{horizon} as a function of  $B_{\rm nG}$.}
		\label{amplitude_dipole8-15_lambda}
\end{figure}

\section{Quantitative predictions for the dipole anisotropy}
\label{sec:results} 

The upper panels of Fig.~\ref{amplitude_dipole8-15} and Fig.~\ref{amplitude_dipole4-8} show the RMS dipole amplitude, $\Delta$, as given by Eq. \ref{dipole}, in the energy bin [8-30] EeV, for different values of the IGMF in the  range  $1<B_{\rm nG}<100$ nG and a field coherence length $\lambda_{\rm Mpc}= 0.2$.  
The power spectrum $P(k)$ is estimated using the CLASS code for a standard $\Lambda$CDM cosmological model \citep{Blas11}.

The different nuclei  display different anisotropies at the same energy. The total anisotropy is $\Delta=\sum X_{i}\Delta_{i}$ where $X_{i}$ is the abundance of nuclei specie $i$. 
As the different nuclei cannot be distinguished by air shower analysis yet,  we simply indicate the expectations for the different components separately.

The anisotropy increases with the magnetic field {(up to $B_{\rm nG}\sim 3$ for CNO, $B_{\rm nG}\sim 10$ for He)}. This is due to the fact that the IGMF impacts both the size of the horizon ({\it i.e.} the distance of the more possible distant source) and the factor $\alpha_1$. 
The anisotropy is larger for a smaller horizon scale, in the case of rectilinear propagation; however after entering the diffusion regime 
the anisotropy is significantly lowered by the factor $\alpha_1$ (since $D \ll rc$ in Eq. \ref{alphagood}).

A larger IGMF coherence length decreases the horizon size and increases the diffusion (decreasing $\alpha_1$), so the two changes would tend to cancel each other out. This can be seen 
in  Fig.~\ref{amplitude_dipole8-15_lambda}, where the resulting dipole for a a coherence length $\lambda_{\rm Mpc}= 1$ is also shown. Note that with  $\lambda_{\rm Mpc}= 1$ the cosmic-ray enter the diffusion regime at smaller values of $B_{\rm nG}$.

In the  $>$ 8 EeV energy bin, anisotropies of the order of the observed one are obtained for a wide range of IGMF values. 
The resulting anisotropy is significantly larger than the one that would result from the Compton-Getting effect \citep{Kachelriess06}. The observed UHECR sources are nearer and hence their fluctuations are larger.

\section{Discussion}
\label{sec:discussion}


If the sources follow the LSS, the RMS dipole moment of the UHECR background is $\Delta\sim 8\%b $, for He and CNO nuclei at energies $\gtrsim8$ EeV, for values  of the IGMF in the range $5\lesssim B_{\rm nG}\lesssim30$ for He and $1\lesssim B_{\rm nG}\lesssim10$ for CNO.  Both ranges are  for a coherence length of 0.2 Mpc. For these value, due to the diffusion in the IGMF the horizon distance, $H\lesssim300$ Mpc, is much smaller than the regular GZK distance at these energies.
A  source bias will increase these values. 
For example, if  UHECR sources follow the galaxy distribution, the bias factor is 
$b\sim1.5$ \citep{2016MNRAS.456.1886S}. 
It might be larger  if the UHECR sources are more clustered than galaxies. 

Our results are consistent with those derived by \citet{2015PhRvD..92f3014H} who found, for a mixed-composition scenario, a dipole amplitude $\sim10$\% at 10 EeV for a source distribution that follows the 2MRS catalog up to $\sim$100 Mpc and a 1 nG IGMF. 
Note, however,  that for this IGMF the magnetic horizon is larger than the extend of the 2MRS catalog, 100 Mpc. Sources at larger distances would lower somewhat the anisotropy.

Unfortunately, these  results cannot be compared directly with the observations  as the UHECRs have to traverse the GMF on their way to Earth.  
The GMF impacts both the amplitude and the direction of the observed dipole \citep{Harari10}.
The GMF has a complex structure \citep{JF12}, and its influence will depend on the direction of the extragalactic dipole. Hence it cannot be studied in a statistical manner. 
Using the \citet{JF12} GMF model, \citet{Taborda17} found that the initial dipole direction (before entering the Galaxy) is 
 in good agreement with the direction of the flux-weighted dipole from the 2MRS galaxy catalog, {\it i.e.} ($l$,~$b$) = (251\degree$\pm$12\degree, 37\degree$\pm$10\degree) for a sample of galaxies within  $\sim$100 Mpc  \citep[][]{Erdogdu06}.  If the UHECR sources are within 100 Mpc and the UHECR dipole is He-dominated, this would imply an IGMF strength $B_{\rm nG}\sim10$. 

Our estimates of  the extragalactic RMS dipole  is $\sim5$\% in the [4-8] EeV range both for protons and He (Fig.~\ref{amplitude_dipole4-8}). The observed anisotropy level in the energy bin [4-8] EeV is significantly lower. If both energy ranges are dominated by the same species ({\it i.e.} He) than this discrepancy can be explained by the propagation in the GMF.

  Another effect that lowers the dipole anisotropy at the [4-8] EeV band is photodisintegration of higher energy nuclei emitted at sources with $d>H$, that we have ignored so far. If the GZK horizon of the daughter nuclei is larger they will reach Earth and contribute to the lower energy bin. As they come from large distances, they would have a lower anisotropy. 
Consider, for example, a  12 EeV C whose daughters are He at 4 EeV. The corresponding GZK distances are $\sim950$ Mpc
and $\sim4000$ Mpc and the magnetic horizons for a 1 nG field are 430 and 885 Mpc.   C emitted from sources between 
the two horizons will photodisintegrate and contribute to He at 4 EeV.
The RMS dipole of these He particles is $\sim1$\% in the quasi-linear regime, $\sim0.1$\% in the diffusive regime.  This can be quantified for a given composition and could significantly reduce the lower band extragalactic dipole. 
Similarly, photodisintegration of He emitted from sources at large distances produces lower energy protons. This lowers the anisotropy and at the same time softens the proton spectrum in the lower energy range, as observed \citep{Apel13}. }

Finally we note that because of different rigidities and different GZK distances, a direct prediction of this model is that different species will show different levels of anisotropy, corresponding to their different rigidities. At present it is impossible to carry out this analysis.  But one can hope that this prediction will be tested in the future.

\begin{acknowledgments}
We acknowledge the I-CORE Program of the Planning and Budgeting Committee and The Israel Science Foundation (grant 1829/12), the advanced ERC grant TReX, and the Lady Davis foundation. 
\end{acknowledgments}


\begin{thebibliography}{}
\bibitem[Aab et al.(2015)]{Aab15} Aab A. [Pierre Auger Collaboration] The Pierre Auger Observatory: Contributions to the 34th International Cosmic Ray Conference, Proc. 34th ICRC, The Hague, The Netherlands
\bibitem[Aab et al.(2017)]{Taborda17}Aab A. [Pierre Auger Collaboration], Science, Vol. 357, Issue 6357, 1266, arXiv:1709.07321
\bibitem[Ade et al.(2013)]{Ade13}P. Ade et al. (Planck Collaboration XXIII), Astron.Astrophys. 571 (2014) A23, arXiv:1303.5083
\bibitem[Amendola et al.(2016)]{2016arXiv160600180A} Amendola, L., Appleby, S., Avgoustidis, A., et al.\ 2016, arXiv:1606.00180 
\bibitem[Apel et al.(2013)]{Apel13} Apel W. et al. (KASCADE-Grande Collaboration),Phys. Rev. D 87 081101 (2013).
\bibitem[Bellido et al.(2017)]{Bellido17}Bellido J. et al. [Pierre Auger Collaboration] Depth of maximum of air-shower profiles at the Pierre Auger Observatory: Measurements above $10^{17.2}$ eV and Composition Implications, 35th ICRC Conference, Jul 2017, Bexco, Busan, Korea, to be published in Proceedings of Science (PoS(ICRC2017)506)
\bibitem[Beck et al.(2016)(2016)]{Beck16}Beck, M. C., Beck, A. M., Beck, R., Dolag K., Strong A. W., \& Nielaba P., 2016, JCAP 5, 056
\bibitem[Blas et al.(2011)]{Blas11}Blas, D., Lesgourgues, J., Tram, T., arXiv:1104.2933 [astro-ph.CO], JCAP 1107 (2011) 034
\bibitem[Davis et al.(2011)]{Davis11}Davis, M., Nusser, A., Masters, K. L., Springob, C., Huchra, J. P., Lemson, G.,  2011, MNRAS, 413, 2906
\bibitem[Deligny et al.(2015)]{Deligny15} The Pierre Auger Collaboration, Astrophys. J. 802 (2015) 111.
\bibitem[Deligny et al.(2015)]{Deligny15} O. Deligny (Pierre Auger and Telescope Array Collaborations) 34th International Cosmic Ray Conference (ICRC 2015), Jul 2015, La Haye, Netherlands. 2015.
\bibitem[Erdo\u{g}du et al.(2006)]{Erdogdu06}Erdo\u{g}du, P., Huchra, J. P., Lahav, O.,  Colless, M., Cutri, R. M., et al., 2006, MNRAS, 368, 1515
\bibitem[Eriksen et al.(2004)]{Eriksen04}H. K. Eriksen, F. K. Hansen, A. J. Banday, K. M. Gorski, and P.B. Lilje, Astrophys. J. 605, 14 (2004); 609, 1198 (2004)
\bibitem[Farrar \& Piran(2000)] {FP00}Farrar, G. R.  \& Piran, T., arXiv:astro-ph/0010370
\bibitem[Ginzburg \& Syrovatskii(1964)]{Ginzburg64}Ginzburg, V. L., \& Syrovatskii, S. I. 1964. The Origin of Cosmic Rays.
\bibitem[Globus et al.(2008)]{GAP08} Globus, N., Allard, D., Parizot, E., 2008, A\&A, 479, 97
\bibitem[Harari et al.(2010)]{Harari10}Harari, Mollerach, Roulet JCAP11 (2010) 033
\bibitem[Harari et al.(2014)]{Harari14}D. Harari, S. Mollerach, E. Roulet, Phys. Rev. D 89, 123001 (2014)
\bibitem[Harari et al.(2015)]{2015PhRvD..92f3014H} Harari, D., Mollerach, S., \& Roulet, E.\ 2015, \prd, 92, 063014 
\bibitem[Jansson \& Farrar (2012)]{JF12}Jansson, R. \& Farrar, G. R., 2012, ApJ, 757, 14
\bibitem[Kachelrie\ss\, \& Serpico(2006)]{Kachelriess06}Kachelrie\ss, M., Serpico, P. D., 2006, Phys. Lett. B 640, 225
\bibitem[Lahav et al.(1997)]{Lahav+97}Lahav, O., Piran, T., \& Treyer, M. A., 1997, MNRAS, 284, 499
\bibitem[Madau et al.(1997)]{Madau97} Madau, P., Meiksin, A., Rees, M. J. 1997, ApJ, 475, 429 
\bibitem[Nusser \& Davis(2011)]{2011ApJ...736...93N} Nusser, A., \& Davis, M.\ 2011, \apj, 736, 93 
\bibitem[Parizot(2004)]{2004NuPhS.136..169P} Parizot, E.\ 2004, Nuclear Physics B Proceedings Supplements, 136, 169 
\bibitem[Pshirkov et al.(2016)]{Pshirkov+16}Pshirkov, M. S., Tinyakov, P. G., Urban F. R., 2016, Phys. Rev. Lett. 116, 191302 
\bibitem[Springob et al.(2016)]{2016MNRAS.456.1886S} Springob, C.~M., Hong, T., Staveley-Smith, L., et al.\ 2016, \mnras, 456, 1886 
\bibitem[Tully et al.(2016)]{2016AJ....152...50T} Tully, R.~B., Courtois, H.~M., \& Sorce, J.~G.\ 2016, \aj, 152, 50 
\bibitem[Unger et al.(2015)]{Unger15} Unger, M., Farrar, G. R. \& Anchordoqui, L. A., 2015, Phys. Rev. D 92, 123001 
\bibitem[Waxman et al.(1997)]{Waxman+97}Waxman, E., Fisher, K. B., Piran, T., 1997, ApJ, 483, 1	
\end{thebibliography}
\end{document}